\documentstyle[amssymb,prl,aps,epsf,floats]{revtex}
\begin{document}
\draft

\narrowtext%
\twocolumn[\hsize\textwidth\columnwidth\hsize\csname @twocolumnfalse\endcsname
\title{Phenomenological description of competing   \\ antiferromagnetism and $d$-wave superconductivity in high $T_{c}$ cuprates}
\author{Bumsoo Kyung and A. M. -S. Tremblay$^1$}
\address{D\'{e}partement de physique and \\ Centre de recherche
sur les propri\'{e}t\'{e}s \'{e}lectroniques
de mat\'{e}riaux avanc\'{e}s.\\ $^1$Institut canadien de recherches avanc\'{e}es\\
Universit\'{e} de Sherbrooke, Sherbrooke, Qu\'{e}bec, Canada J1K 2R1}
\date{April 23, 2002}
\maketitle
\begin{abstract}
In this paper the phase diagram
of high $T_{c}$ cuprates is {\it qualitatively} studied in the context of
competing orders: antiferromagnetism, $d$-wave superconductivity
and $d$-density wave.
{\it Local} correlation functions are estimated from a mean-field solution of
the $t-J$ Hamiltonian. With decreasing doping
the superconducting mean-field $T^{MF}_{c}$ and order parameter 
$d$ begin to decrease
below some characteristic doping $x_{c} \simeq 0.2$ where
short-range antiferromagnetic correlations begin to develop.
{\it Dynamical} properties that involve the energy spectrum,
such as the normal state pseudogap, are calculated from effective interactions
that are consistent with the above-mentioned local correlation functions.
The total excitation gap $\Delta_{tg}$ (in the superconducting state) and
the normal state pseudogap $\Delta_{pg}$
are in good agreement with experimental results.
Properties of the condensate are estimated using 
an effective pairing interaction
$V_{eff}$ which takes into account (pair breaking) 
antiferromagnetic correlations.
These condensate properties include condensation  energy $U(0)$,
coherence gap $\Delta_{cg}$ and critical field $H_{c2}$.
The calculated coherence gap closely follows the doping
dependence of $T_{c}$ or $d$, and is approximately given
as $\Delta_{cg} \sim \Delta_{tg}-\Delta_{pg}$
within our numerical uncertainties.
The systematic decrease of superfluidity ($d$, $U(0)$, $\Delta_{cg}$, $H_{c2}$),
and systematic increase of
$\Delta_{pg}$ and $\Delta_{tg}$ with decreasing doping below $x_{c}$
have their natural explanation in our approach.  
The overall description is however 
qualitative since it does not appear possible
to obtain results that are in quantitative agreement with experiment
for all physical quantities. 
\end{abstract}
\vskip2pc] \narrowtext

\section{Introduction}

\label{section1}

Fifteen years have passed since the discovery of high temperature
superconductivity in Ln$_{2-x}$Ba$_{x}$CuO$_{4}$ by Bednorz and M\"{u}ller, %
\cite{Bednorz:1986} yet there is no widely accepted theoretical explanation
of the many anomalous features found in these materials. This is due to a
large extent to the lack of reliable theoretical methods to study this
strongly-correlated electron problem. Dynamical Mean Field theory, for
example,\cite{Georges:1996} works extremely well in three or higher
dimensions but it lacks the momentum-dependent self-energy effects that are
intrinsic to a two-dimensional situation with strong antiferromagnetic
fluctuations. More recent dynamical cluster approaches correct this defect
but these methods are still in their infancy and are entirely numerical.\cite%
{Jarrell:2000}\cite{Kotliar:2001} In this paper, we step back and take a
more humble stand, seeking the {\it simplest} {\it phenomenological}
approach that describes the main {\it overall }features of high-temperature
superconductors. Let us recall what these features are.

First consider the generic phase diagram of a hole-doped cuprate Ln$_{2-x}$Sr%
$_{x}$CuO$_{4}$ in the doping ($x=1-n$) and temperature ($T$) plane.\cite%
{Almasan:1991} Near half-filling and at low temperature, antiferromagnetic
(AF) long-range order appears and is destroyed by $2\%$ doping concentration
($x=0.02$). When $x$ reaches $0.05$, superconducting (SC) long-range order
starts to appear and is also destroyed by about $30\%$ doping. In between
them, $T_{c}$ reaches a maximum value at $x\simeq 0.15$. The SC gap was
found to have mainly $d$-wave character,\cite%
{Hardy:1993,Levi:1993,Shen:1993,Marshall:1996} in contrast to conventional
BCS superconductors\cite{Bardeen:1957} with an isotropic $s$-wave gap.

Second, various recent experiments also show the existence of a crossover
temperature $T^{\ast }$ higher than $T_{c}$ in underdoped and optimally
doped samples. Below this so-called pseudogap temperature $T^{\ast }$,
low-frequency spectral weight begins to be strongly suppressed in many
physical quantities. This anomalous phenomenon has been observed in angle
resolved photoemission spectroscopy (ARPES),\cite{Ding:1996,Loeser:1996}
specific heat,\cite{Loram:1993} tunneling,\cite{Renner:1998} NMR,\cite%
{Takigawa:1991} and optical conductivity.\cite{Homes:1993} Surprisingly the
doping dependences of $T^{\ast }$ and $T_{c}$ are completely different\cite%
{Timusk:1999,Tallon:2001} in spite of their possibly close relationship
suggested by ARPES\cite{Ding:1996,Loeser:1996} and tunneling\cite%
{Renner:1998} experiments. At optimal doping, where $T_{c}$ is maximum,
various non-Fermi liquid (NFL) properties are observed in the normal state.
These include the linear temperature dependence (up to $1000K$) of ab-plane
resistivity, the quadratic $T$ dependence of Hall angle and so on. Far
beyond optimal doping, there are indications that the normal state
properties are probably well described by the conventional Landau Fermi
liquid. Near and below optimal doping, the superfluid density $n_{s}/m$ is
systematically suppressed with decreasing doping in spite of the increasing
excitation gap amplitude, as measured by ARPES, in the superconducting
state. The resulting $2\Delta _{max}/k_{B}T_{c}$ ratio is very different
from the universal BCS value, $4.3$ for $d$-wave symmetry. In the overdoped
regime, however, the SC properties are probably well explained in the
conventional weak-coupling BCS theory, although this has not yet been
checked in detail.\cite{Taillefer:2001}

The approach that we take is as follows. First, we do a simple mean-field
analysis of the $t-J$ Hamiltonian whose results are used only to obtain
estimates for zero-temperature order parameters and to compute equal-time
local correlation functions (such as double occupancy). Dynamical
quantities, such as the response functions and single-particle spectral
weight, are computed by using the values of the local correlation functions
obtained in mean field to fix the value of the renormalized interaction
vertices. We do not consider the possibility of spin-charge separation\cite%
{Anderson:2000,Senthil:2000} but we take into account the
no-double-occupancy constraint by replacing projectors in the $t-J$
Hamiltonian by their mean-field bosonic value, namely doping $x$. This is an
effective-mass approximation. We cannot address the issue of a quantum
critical point (QCP) \cite{Sachdev:1999} at $x_{c}$ even if it exists at the
mean-field level. Non-Fermi liquid behavior above the pseudogap temperature
is also beyond the domain of validity of our approach. The regime we
consider is one where the temperature would normally be low enough for a
Fermi liquid to appear but where fluctuations destroy that phase, especially
in the underdoped regime.

The paper is organized as follows: In Section~\ref{section2} the $t-J$
Hamiltonian is decoupled, in a mean-field approximation, with competing (SC,
AF and $d-$density wave) order parameters. The mean-field Hamiltonian is
diagonalized to compute mean-field ordering temperatures $T_{c}^{MF}$, $%
T_{N}^{MF}$ and mean-field order parameters $d$ and $m$. Dynamical
quantities such as the normal-state pseudogap are computed and compared with
experimental results in Section~\ref{section4}. In Section~\ref{section5}
the results of the previous sections are gathered to discuss the phase
diagram of high $T_{c}$ cuprates. In Section~\ref{sectionSC} we discuss some
of the properties of the SC condensate, such as the condensation energy,
upper critical magnetic field and the coherence gap, using a model that
takes into account the effect of (pair breaking) AF correlations in an
effective manner. The present results are compared with some of the leading
theories for the high $T_{c}$ superconductivity in Section~\ref{section6} as
well as with experiments. The last Section summarizes the results and
discusses its main limitation.

\section{Mean-field $t-J$ Hamiltonian}

\label{section2}

Right after the discovery of high temperature superconductors, Anderson\cite%
{Anderson:1987} first proposed the one-band Hubbard model as the simplest
Hamiltonian that might capture the correct low energy physics of copper
oxides. The $t-J$ model is known to be the large $U$ limit of the Hubbard
Hamiltonian under certain assumptions. The $t-J$ model is described by the
Hamiltonian 
\begin{eqnarray}
H &=&-t\sum_{\langle i,j\rangle ,\sigma }\bigl((1-n_{i,-\sigma })c_{i,\sigma
}^{\dag }c_{j,\sigma }(1-n_{j,-\sigma })+\mbox{H.c.}\bigr)  \nonumber \\
&+J&\sum_{\langle i,j\rangle }\bigl(\vec{S}_{i}\cdot \vec{S}_{j}-\frac{1}{4}%
n_{i}n_{j}\bigr)-\mu \sum_{i,\sigma }c_{i,\sigma }^{\dag }c_{i,\sigma }\;,
\label{eq2-10}
\end{eqnarray}%
where $c_{i,\sigma }$ destroys an electron at site $i$ with spin $\sigma $
on a two-dimensional square lattice, $t$ is the hopping matrix element
between nearest neighbors $<i,j>$ and $J$ denotes superexchange coupling.
Double occupancy of two electrons at the same lattice site is forbidden in
the Hilbert space corresponding to Eq.(\ref{eq2-10}) as can be seen by the
projection operator $(1-n_{i,-\sigma })$ in the hopping term. The chemical
potential $\mu $ controls the electron density $n$. Finally, $\vec{S}_{i}$
and $n_{i}$ are spin and charge density operators, respectively, and they
are defined by 
\begin{eqnarray}
\vec{S}_{i} &=&\frac{1}{2}\sum_{\alpha ,\beta }c_{i,\alpha }^{\dag }\vec{%
\sigma}_{\alpha ,\beta }c_{i,\beta }\;,  \nonumber \\
n_{i} &=&\sum_{\sigma }c_{i,\sigma }^{\dag }c_{i,\sigma }\;,  \label{eq2-20}
\end{eqnarray}%
where the components of $\vec{\sigma}$ are $2\times 2$ spin Pauli matrices.

In general the (static) mean-field description is not accurate for strongly
correlated electron systems such as the Hubbard and $t-J$ models in the
physically relevant regime. Nevertheless it is known that, for example, the
mean-field description of the antiferromagnetic state at $x=0$ is quite
accurate, even at large $U.$ For example, spin waves with accurate spin-wave
velocities can be obtained by performing $RPA$ in the ordered phase.\cite%
{Schrieffer:1989} We take the point of view that mean-field theory can be
useful to select the leading correlations when there are several {\it %
competing} order parameters in the low energy sector. Due to the neglect of
spatial and quantum fluctuations in mean-field theory, some caution is
necessary in interpreting the results. First, the mean-field phase line is
to be decreased in temperature to take into account the effect of quantum
fluctuations (such as Kanamori-Brueckner screening\cite{Vilk:1997}). Second,
since we work in two dimensions where antiferromagnetic long-range order is
prohibited by the Mermin-Wagner theorem, the renormalized mean-field N\'{e}%
el transition line has to be interpreted as the onset of the corresponding 
{\it short-range} correlations instead of as a {\it true} thermodynamic
phase transition. Whether given short-range correlations eventually grow to
long-range order at low temperature or not, should be answered by studying
how the zero-frequency correlation length grows when temperature is lowered.
This question, as well as that of dynamical or quantum fluctuations that may
take place on very short distance scales\cite{Georges:1996} can be examined
by the fluctuation theory that we describe in Section \ref{section4}.

In the spirit of a mean-field approximation, terms with more than two
operators should be decoupled in all possible ways. In principle there are
infinitely many ways of decoupling the Heisenberg part of the Hamiltonian.
In this situation, guidance from experiments and past theoretical
considerations is helpful to find the most important leading correlations of
the model, correlations whose stability is studied {\it a posteriori}. We
consider three {\it competing} mean-field order parameters, $m$, $d$, and $y$
for AF, $d$-wave SC, and $d$-density wave\cite{Chakravarty:2001} orders
respectively, which are defined\cite{NoteFactor} by 
\begin{eqnarray}
m=\langle \hat{m}\rangle &=&1/(2N)\sum_{\vec{k},\sigma }\sigma \langle c_{%
\vec{k}+\vec{Q},\sigma }^{\dag }c_{\vec{k},\sigma }\rangle , \\
d=\langle \hat{d}\rangle &=&1/N\sum_{\vec{k}}\phi _{d}(\vec{k})\langle c_{%
\vec{k},\uparrow }c_{-\vec{k},\downarrow }\rangle \;, \\
y &=&\langle \widehat{y}\rangle =i/\left( 2N\right) \sum_{\vec{k},\sigma
}\phi _{d}(\vec{k})\langle c_{\vec{k}+\vec{Q},\sigma }^{\dag }c_{\vec{k}%
,\sigma }\rangle ,
\end{eqnarray}%
where $N$ is the total number of lattice sites, $\phi _{d}(\vec{k})=\cos
k_{x}-\cos k_{y}$ is the $d-wave$ form factor and $\vec{Q}$ is the
(commensurate) AF wave vector $(\pi ,\pi )$ in two dimensions. The extended $%
s-$wave analogs of $d-$wave superconductivity and $d-$density wave orders
are also present in the $t-J$ Hamiltonian, but they are always less relevant
in mean-field studies. In this paper we restrict ourselves to a uniform
solution. At the end of the paper the issue of inhomogeneous modulation of
spin and charge degrees of freedom will be briefly discussed in the context
of the present results. A similar mean-field decoupling for the AF and SC
channels only was previously considered by several groups.\cite%
{Inui:1988,Inaba:1996} In the present study the projected operator $%
c_{j,\sigma }(1-n_{j,-\sigma })$ is separated into the original electron
operator $c_{i,\sigma }$ times the expectation of the bosonic operator $%
(1-n_{j,-\sigma })$ which is taken as $x^{1/2}$. As long as the
no-double-occupancy constraint is {\it globally} imposed, the average number
of electron $n_{e}$ in our study is equal to $1-n_{h}$ with $x=n_{h}$.
Clearly, at half-filling, one recovers the Heisenberg Hamiltonian. In
addition, as we shall see later, for $T=0$, $n=1,$ the no-double-occupancy
constraint is satisfied {\it exactly }because the AF solution allows only
one electron per site with the full moment. This would not be the case if we
considered only $d-$wave and $d-$ density wave mean-field solutions.

For a more systematic implementation of this constraint beyond the
mean-field level, the slave boson representation can be useful.\cite%
{Lee:2000,Nayak:2000} In the slave boson representation, an electron is
explicitly decomposed into a spinon (fermion) and a holon (boson), $%
c_{i,\sigma }^{+}=f_{i,\sigma }^{+}b_{i}$. The latter bosonic operator
corresponds to $(1-n_{i-\sigma })$ in our approach and our global constraint 
$1=n_{e}+n_{h}$ corresponds to the no double-occupancy constraint in slave
boson mean-field theories.

In terms of mean-field order parameters $m$, $d$, and $y,$ the mean-field $%
t-J$ Hamiltonian reads 
\begin{equation}
H_{MF}=\sum_{\vec{k},\sigma }\varepsilon (\vec{k})c_{\vec{k},\sigma }^{\dag
}c_{\vec{k},\sigma }-4Jm\hat{m}-Jd(\hat{d}+\hat{d}^{\dag })-Jy\widehat{y}%
+F_{0}\;,  \label{eq2-40}
\end{equation}%
where 
\begin{equation}
F_{0}=N(2Jm^{2}+Jd^{2}+\frac{J}{2}y^{2}-\mu )\;.  \label{F0}
\end{equation}%
$\varepsilon (\vec{k})\simeq -2tx(\cos k_{x}+\cos k_{y})-\mu $ with $x$ the
hole density. The correlated hopping is taken into account through $x$ in a
mean-field spirit, as explained above. In other words, the
no-double-occupancy constraint manifests itself here as an effective mass,
not as a change in the number of carriers. One has to keep in mind however
that, at this level, $\varepsilon (\vec{k})$ and the corresponding effective
mass are only bare values that are renormalized by interactions. This is
discussed further at the end of the following section.

Previous studies have suggested that the $d-$density wave order parameter
becomes important only quite close to half-filling.\cite{Ubbens:1992} Here
we find that this order parameter is {\it never} a self-consistent solution
of the mean-field equations when mean-field AF order is explicitly
considered in Eq.(\ref{eq2-40}). Hence, in the following, we restrict
ourselves to AF and SC order parameters. By introducing a four component
field operator $\Psi _{\vec{k}}^{\dag }$ 
\[
\Psi _{\vec{k}}^{\dag }=(c_{\vec{k},\uparrow }^{\dag },c_{-\vec{k}%
,\downarrow },c_{\vec{k}+\vec{Q},\uparrow }^{\dag },c_{-\vec{k}-\vec{Q}%
,\downarrow })\;, 
\]%
Eq.~(\ref{eq2-40}) may be written in a more compact form to reflect the
competition between AF and SC order, 
\[
H_{MF}=\sum_{\vec{k}}{}^{^{\prime }}\Psi _{\vec{k}}^{\dag }M_{\vec{k}}\Psi _{%
\vec{k}}+F_{0}\;. 
\]%
The prime symbol on the summation restricts the summation over wave vectors
to the magnetic Brillouin zone in order to take into account the doubling of
the unit cell in the presence of (commensurate) AF order. Small
incommensuration would not change appreciably the value of the local
correlation functions that we will compute with the mean-field solution,
hence we limit ourselves to the commensurate case. The matrix $M_{\vec{k}}$
in the last equation is given by 
\begin{eqnarray}
M_{\vec{k}} &=&\left(
\begin{array}{cccc}
\varepsilon (\vec{k}) & -Jd\phi _{d}(\vec{k}) & -2Jm & 0 \\
-Jd\phi _{d}(\vec{k}) & -\varepsilon (\vec{k}) & 0 & -2Jm \\
-2Jm & 0 & \varepsilon (\vec{k}+\vec{Q}) & Jd\phi _{d}(\vec{k}) \\
0 & -2Jm & Jd\phi _{d}(\vec{k}) & -\varepsilon (\vec{k}+\vec{Q})
\end{array}
\right) \;.  \nonumber  \\
&&
%M_{\vec{k}} &=&\left( 
%\begin{array}{cccc}
%\varepsilon (\vec{k}) & -Jd\phi _{d}(\vec{k}) & -2Jm & 0 \\ 
%-Jd\phi _{d}(\vec{k}) & -\varepsilon (\vec{k}) & 0 & -2Jm \\ 
%-2Jm & 0 & \varepsilon (\vec{k}+\vec{Q}) & Jd\phi _{d}(\vec{k}) \\ 
%0 & -2Jm & Jd\phi _{d}(\vec{k}) & -\varepsilon (\vec{k}+\vec{Q})%
%\end{array}%
%\right) \;. \\
%&&
\end{eqnarray}%
The energy eigenvalues of $M_{\vec{k}}$ yield four branches to the energy
dispersion $\pm E_{\pm }(\vec{k})$ 
\begin{eqnarray}
E_{\pm }(\vec{k}) &=&[(\varepsilon _{\vec{k}}^{2}+\varepsilon _{\vec{k}+\vec{%
Q}}^{2})/2+(2Jm)^{2}+(Jd\phi _{d}(\vec{k}))^{2}  \nonumber \\
&&\pm g(\vec{k})\;\;]^{1/2}\;,  \label{eq2-80}
\end{eqnarray}%
where $g(\vec{k})$ is given as 
\begin{equation}
g(\vec{k})=[(\varepsilon _{\vec{k}}^{2}-\varepsilon _{\vec{k}+\vec{Q}%
}^{2})^{2}/4+((\varepsilon _{\vec{k}}+\varepsilon _{\vec{k}+\vec{Q}%
})(2Jm)]^{1/2}\;.
\end{equation}

The free energy is easily obtained either from the trace formula or from the
Feynman theorem 
\[
F=-2T\sum_{\vec{k}}^{^{\prime }}\sum_{\alpha =\pm }\log (2\cosh \frac{%
E_{\alpha }(\vec{k})}{2T})+F_{0}\;. 
\]%
Two mean-field equations are obtained from the stationary condition on $F$
with respect to the corresponding order parameters, $\frac{\partial F}{%
\partial m}=\frac{\partial F}{\partial d}=0$, and one more unknown constant $%
\mu $ is determined by the thermodynamic relation $n=1-x=-\frac{\partial F}{%
\partial \mu }$. The resulting three equations are 
\begin{eqnarray}
&&m=\frac{1}{2N}\sum_{\vec{k}}^{^{\prime }}\sum_{\alpha =\pm
}\Bigl\{(2Jm)+\alpha \frac{(\varepsilon _{\vec{k}}+\varepsilon _{\vec{k}+%
\vec{Q}})^{2}(2Jm)}{2g(\vec{k})}\Bigr\}  \nonumber  \label{eq2-110} \\
&&\times \frac{1}{E_{\alpha }(\vec{k})}\tanh (\frac{\beta E_{\alpha }(\vec{k}%
)}{2})\;,  \nonumber \\
&&d=\frac{1}{2N}\sum_{\vec{k}}^{^{\prime }}\sum_{\alpha =\pm }\phi _{d}^{2}(%
\vec{k})(Jd)\frac{1}{E_{\alpha }(\vec{k})}\tanh (\frac{\beta E_{\alpha }(%
\vec{k})}{2})\;,  \nonumber \\
&&n=1-\frac{1}{2N}\sum_{\vec{k}}^{^{\prime }}\sum_{\alpha =\pm
}\Bigl\{(\varepsilon _{\vec{k}}+\varepsilon _{\vec{k}+\vec{Q}})  \nonumber \\
&&+\alpha \frac{(\varepsilon _{\vec{k}}+\varepsilon _{\vec{k}+\vec{Q}%
})(\varepsilon _{\vec{k}}-\varepsilon _{\vec{k}+\vec{Q}})^{2}}{2g(\vec{k})}
\nonumber \\
&&+\alpha \frac{2(2Jm)^{2}(\varepsilon _{\vec{k}}+\varepsilon _{\vec{k}+\vec{%
Q}})}{g(\vec{k})}\Bigr\}\frac{1}{E_{\alpha }(\vec{k})}\tanh (\frac{\beta
E_{\alpha }(\vec{k})}{2})\;.  \nonumber \\
&&
%&&m=\frac{1}{2N}\sum_{\vec{k}}^{^{\prime }}\sum_{\alpha =\pm
%}\Bigl\{(2Jm)+\alpha \frac{(\varepsilon _{\vec{k}}+\varepsilon _{\vec{k}+%
%\vec{Q}})^{2}(2Jm)}{2g(\vec{k})}\Bigr\} \\
%&&\times \frac{1}{E_{\alpha }(\vec{k})}\tanh (\frac{\beta E_{\alpha }(\vec{k}%
%)}{2})\;,  \nonumber \\
%&&d=\frac{1}{2N}\sum_{\vec{k}}^{^{\prime }}\sum_{\alpha =\pm }\phi _{d}^{2}(%
%\vec{k})(Jd)\frac{1}{E_{\alpha }(\vec{k})}\tanh (\frac{\beta E_{\alpha }(%
%\vec{k})}{2})\;,  \nonumber \\
%&&n=1-\frac{1}{2N}\sum_{\vec{k}}^{^{\prime }}\sum_{\alpha =\pm
%}\Bigl\{(\varepsilon _{\vec{k}}+\varepsilon _{\vec{k}+\vec{Q}})  \nonumber \\
%&&+\alpha \frac{(\varepsilon _{\vec{k}}+\varepsilon _{\vec{k}+\vec{Q}%
%})(\varepsilon _{\vec{k}}-\varepsilon _{\vec{k}+\vec{Q}})^{2}}{2g(\vec{k})} 
%\nonumber \\
%&&+\alpha \frac{2(2Jm)^{2}(\varepsilon _{\vec{k}}+\varepsilon _{\vec{k}+\vec{%
%Q}})}{g(\vec{k})}\Bigr\}\frac{1}{E_{\alpha }(\vec{k})}\tanh (\frac{\beta
%E_{\alpha }(\vec{k})}{2})\;.
\end{eqnarray}%
The spin-triplet order parameter $\langle c_{\vec{k}+\vec{Q},\uparrow }c_{-%
\vec{k},\downarrow }\rangle $ is ``dynamically generated''\cite{Kyung:2000-1}
when the two mean-field order parameters $m$ and $d$
coexist, even when it is not explicitly included in the mean-field
factorization. We have checked that including this order parameter in the
mean-field factorization changes the value of local correlation functions,
such as $\langle |\Delta _{d}(0)|^{2}\rangle $ which we need below for the
dynamical calculation, by less than one percent. The values of $m$ and $d$,
on the other hand, change by about $10\%$. The inclusion of this order
parameter would then change the results of the present paper only
quantitatively and at the few percent level only. We do not include it for
now for simplicity. It should be included at a later stage when refining the
approach.

\begin{figure}
\centerline{\epsfxsize 8cm \epsffile{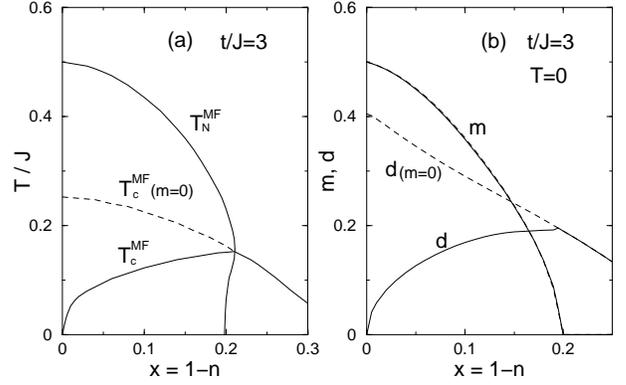}}
\caption{(a) Calculated mean-field phase diagram in doping ($x=1-n$)
and temperature ($T$) plane and (b) mean-field order
parameters for $t/J=3$.
$T^{MF}_{N}$ and $T^{MF}_{c}$ are mean-field AF and SC
ordering temperatures,
while $m$ and $d$ are mean-field AF and SC order parameters.
$T^{MF}_{N} (m=0)$ and $d (m=0)$ (dashed curves) are
SC ordering temperature
and order parameter, respectively, when the mean-field factorization
of the interaction term (J) is in the SC channel only.
The dashed curve for $T^{MF}_{N}(d=0)$ is also shown in (b) but it is
indistinguishible from the case where $d$ takes its mean-field value.}
\label{fig1}
\end{figure}%
Even though it is only an intermediate step in our calculations, for
completeness we show in Fig.\ref{fig1} the calculated mean-field phase
diagram and zero-temperature mean-field order parameters $m$ and $d$ for $%
t/J=3$. The AF order parameter dominates near half-filling and vanishes
beyond $x=x_{c}\simeq 0.20$. On the other hand, $d$-wave SC correlations
keep growing with decreasing doping down to $x_{c}\simeq 0.20$ where the AF
order parameter is non-zero. Below $x_{c}$, we observe that $T_{c}^{MF}$ and 
$d$ begin to monotonically decrease with decreasing doping. The physical
origin of this decrease of $T_{c}^{MF}$ and $d$ should be a combination of
the effects of local moment formation and of the onset of AF correlations
that break time-reversal symmetry, both of which lead to the destruction of
the SC state. The importance of AF correlations in reducing $T_{c}$ to the
experimental value ($\sim $ $100K$) was recently pointed out by Weng.\cite%
{Weng:1999} In between half-filling and $x_{c}$, AF and SC orders coexist at
the \thinspace mean-field level. The overall features are similar to
previous results\cite{Kyung:2000-1} based on a phenomenological model in
which SC and AF orders come from different interaction terms.

The dashed curves in Fig.\ref{fig1}(a) and (b) are $T_{c}^{MF}$ and $d$
when the interaction piece of the Hamiltonian is decoupled only in the
(spinon) pairing channel in the slave boson mean-field theory\cite%
{Baskaran:1987} or, equivalently, when $m$ is forced to zero in the present
approach. This shows that, by itself, the diverging effective mass is not
enough to reduce $T_{c}^{MF}$ and $d$ to zero at half-filling. In the
slave-boson theory, $T_{c}^{MF}$ and $d$ keep increasing with decreasing
doping. $T_{c}^{MF}$ is interpreted as pseudogap temperature $T^{\ast }$ and
the corresponding ground state is dubbed as the resonating valence bond
(RVB) state.\cite{Anderson:1987} In this theory, the true $T_{c}$ is
governed by the Bose-condensation temperature of holons, which scales as $x$%
. Then the maximum $T_{c}$ occurs at very small doping concentration ($%
x<0.05 $). More recent versions of this slave-boson mean-field theory seem
to correct this defect\cite{Brinckmann:2001}.

In the present study, the behavior of the mean-field order parameters
reflects the competition between AF and SC correlations. As can be seen in
Fig.\ref{fig1}, SC correlations are most significantly modified due to the
presence of (pair breaking) AF correlations, while the latter are virtually
unchanged by the former. When Eq.~\ref{eq2-40} is converted into an
effective Hamiltonian where the competing order parameters are expressed in
terms of operators we find 
\begin{equation}
H=\sum_{\vec{k},\sigma }\varepsilon (\vec{k})c_{\vec{k},\sigma }^{\dag }c_{%
\vec{k},\sigma }-2J\hat{m}\hat{m}-J\hat{d}^{\dag }\hat{d}-\frac{1}{2}J%
\widehat{y}\widehat{y}.  \label{MF1}
\end{equation}%
In this notation, it is clear that interactions that generate AF are
stronger than those that generate SC by a factor of two and that the
interactions that generate $d-$density wave order are even smaller. Note
that to be able to make this comparison, the order parameters are normalized
\cite{NoteFactor}
such that the largest value that either one of them can take in a mean-field
solution is near $1/2$ (when all other order parameters are forced to zero).
In particular then, even if we drop the $d-$density wave, this mean-field $%
t-J$ Hamiltonian is not $SO(5)$ symmetric\cite{Zhang:1997} since in that
case, the $\hat{m}\hat{m}$ and $\hat{d}^{\dag }\hat{d}$ terms would have the
same coefficients, so as to form a five-dimensional superspin vector.\cite%
{NoteDefSO(5)}

\section{Dynamical properties : excitation pseudogaps}

\label{section4}

In a previous study of the two-dimensional attractive Hubbard model\cite%
{Kyung:2001-3} it was shown, by comparisons with Quantum Monte Carlo
calculations, that the BCS mean-field ground state gave a good estimate (few
percent) of double-occupancy in the normal state when the interaction
strength satisfied $|U|\gtrsim 3t$ and when the temperature is reasonably
low ($T$ of order $0.5t$ or less). We expect this should remain true for
correlation functions that are local in space and time. On the other hand,
for dynamical quantities mean-field theory itself is clearly a bad
estimate. Instead of a true antiferromagnetic gap as in Fig.\ref{fig1}, we
expect to have a pseudogap.

In Ref.\cite{Kyung:2001-4}, the procedure used to compute the pseudogap size 
$\Delta _{pg}$ was as follows. Let $\Delta _{d}(i)$ be defined by 
\[
\Delta _{d}(i)=\frac{1}{2}\sum_{\delta }g(\delta )(c_{i+\delta ,\uparrow
}c_{i,\downarrow }-c_{i+\delta ,\downarrow }c_{i,\uparrow })\;,
\]%
where $g(\delta )$ is a $d$-wave form factor given, in real space
representation, by $g(\delta )=1/2$ for $\delta =(\pm 1,0)$, $-1/2$ for $%
\delta =(0,\pm 1)$, and $0$ otherwise. The $d$-wave symmetry spin-singlet
local correlation function, $\langle |\Delta _{d}(0)|^{2}\rangle $, is
computed from the ground state mean-field solution \cite{Kyung:2001-4} but
with only $m$ different from zero. In other words, it contains only AF
correlations. This correlation function, $\langle |\Delta
_{d}(0)|^{2}\rangle $, is then used in the fluctuation-dissipation theorem
to obtain a renormalized vertex that allows one to estimate dynamical
susceptibilities in any channel.\cite{Note-1} 
The latter come in the computation of both
the two-particle and one-particle quantities. An example of such a
calculation is given in section\ref{sectionSC}. The pseudogap appears in the
renormalized classical regime\cite{Vilk:1996} of the singlet $d$-wave
fluctuations in a region of the Fermi surface, namely the zone edges, where
the thermal de Broglie wavelength is so small that it can become smaller
than the $d$-wave zero-frequency correlation length even if the latter is of
order only one lattice spacing. To be more specific, for $x=0.1$ for
example, the pseudogap appears near the zone edge when $T=0.40J$. At this
temperature, the characteristic energy of the $d-$wave singlet fluctuations
is $0.31J$ while the corresponding correlation length estimated from the
wave vector dependence of the $\omega =0$ $d-$wave singlet susceptibility
around $q=0$ is $0.48$ lattice spacing.

\begin{figure}
\centerline{\epsfxsize 8cm \epsffile{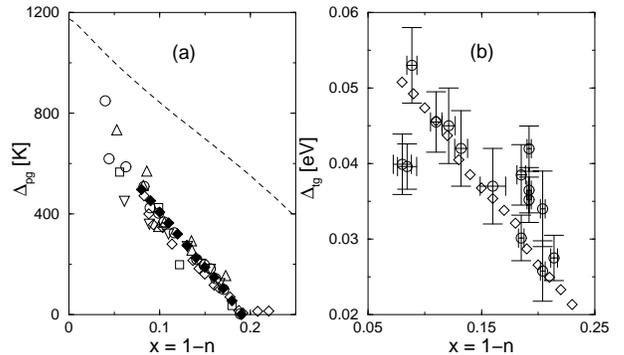}}
\caption{(a) Calculated pseudogap size at $T=0$, $\Delta_{pg}$,
denoted as filled diamonds for $t/J=3.0$ and $J=125$ meV.
The open circles, diamonds, squares, up-triangles, and down-triangles
are the pseudogap size extracted from susceptibility,
heat capacity, ARPES, NMR, and resistivity measurements, respectively,
by Tallon and Loram.\protect\cite{Tallon:2001}
The dashed curve denotes (spinon) pseudogap size predicted by the
slave boson mean-field theory of the $t-J$ model.
(b) Calculated total excitation gap at $T=0$, $\Delta_{tg}$
(empty diamonds). The empty circles with error bars denote
ARPES leading edge gap by Ding {\it et al.}
\protect\cite{Ding:2000}}
\label{fig3}
\end{figure}%
In Fig.~\ref{fig3}(a) the calculated pseudogap size $\Delta _{pg}$ (filled
diamonds) from Ref.\cite{Kyung:2001-4} is plotted along with the pseudogap
energy (empty symbols) extracted from various experiments by Tallon and
Loram.\cite{Tallon:2001} Note that the pseudogap rapidly takes its
asymptotic low-temperature value when it opens up at finite temperature.
That is the reason why we plot the value it would have at $T=0$ if the
superconducting phase did not appear. Since in the above calculational
procedure, the local $d-$wave singlet correlation function $\langle |\Delta
_{d}(0)|^{2}\rangle $ that is used to compute the dynamical singlet
susceptibility is computed from static antiferromagnetic mean-field order,
the corresponding normal state pseudogap $\Delta _{pg}$ vanishes at $x\simeq
x_{c}$. The calculated $\Delta _{pg}$ is in good agreement with experimental
results.\cite{NoteApprox} Through interlayer tunneling spectroscopy of $%
Bi_{2}Sr_{2}CaCu_{2}O_{8+\delta }$ Krasnov\cite{Krasnov:2002} also observed
that the pseudogap decreases approximately linearly with doping and vanishes
at the critical doping, $x_{c}\simeq 0.19$.

The dashed curve is the pseudogap size predicted from the slave-boson
mean-field theory of the $t-J$ Hamiltonian. \cite{Baskaran:1987} In fact the
(spinon) pseudogap size obtained in the latter theory is the same as $2Jd$
calculated from Eq.(\ref{eq2-40}) with $m=0$. It is much larger than the
experimentally extracted pseudogap energy throughout the interesting doping
range. Furthermore there is no indication that it vanishes near $x_{c}$, in
conflict with experimental results. Although spin degrees of freedom are the
main driving force for the pseudogap behavior in the above theory and in
ours, the detailed implementation is qualitatively different. In the
slave-boson mean-field theory a pseudogap is caused by a pairing among {\it %
spinons}, while in our theory it is {\it short-range} spin correlations of 
{\it electrons} that induce an effective attraction in the particle-particle
channel. The spinon pseudogap in the former theory has nothing to do with
local moment formation and their short-range AF correlations with decreasing
temperature.

The total excitation gap $\Delta _{tg}$, at $T=0$, is obtained\cite%
{Kyung:2001-4} by allowing both $d$ and $m$ to be different from zero in the
calculation of the quantity $\langle |\Delta _{d}(0)|^{2}\rangle $ that is
used in the fluctuation-dissipation theorem to obtain the renormalized
vertices entering the calculation of dynamical quantities. The calculated
total excitation gap (or ARPES leading edge gap or SC gap), $\Delta _{tg}$,
at $T=0$ is also plotted in Fig.~\ref{fig3}(b) as a function of doping with $%
t/J=3$. The open circles with error bars are experimentally determined
leading edge gap by Ding {\it et al.}\cite{Ding:2000} in their ARPES
measurement. Except a few points, which may be due to the effect of
impurities as noted by the authors, our calculated $\Delta _{tg}$ is in
reasonable agreement with experimental results and decreases more or less
linearly with doping. In addition, the velocity defined by 
\[
v_{2}=\frac{\Delta _{tg}}{\sqrt{2}}\sin \frac{k_{F}}{\sqrt{2}} 
\]%
ranges from $1.0$ to $2.0\times 10^{6}cm/s$ when $x$ varies from $0.2$ to $%
0.1.$ Experiments on various compounds and dopings in that range give
results of order $1.0\times 10^{6}cm/s$ according to Ref.\cite{Chen:2000}
More recent ARPES results Bi-2212\cite{Mesot:1999} give a ratio of $%
v_{2}/\Delta _{tg}$ that can decrease by about a factor two from optimal to
underdoping, while transport measurements suggest a much stronger doping
dependence.\cite{Taillefer:2001} Note that in our calculations of $\Delta
_{pg}$ and $\Delta _{tg}$ we use the experimentally determined value of
exchange coupling $J=125$ meV. The only adjustable parameter is the ratio $%
t/J$ that we take, as in the previous sections, equal to $3$.

$\Delta _{tg}$ is always larger than $\Delta _{pg}$ due to the additional
contribution from $d\neq 0$\cite{Kyung:2001-4} to the local spin-singlet
amplitude $\langle |\Delta _{d}(0)|^{2}\rangle $. Since the SC order
parameter vanishes at $T_{c}$, the SC gap below $T_{c}$ continuously evolves
into the normal state pseudogap above $T_{c}$ with the same momentum
dependence and magnitude.

Finally we comment on the Fermi velocity in the nodal direction. \cite%
{Mesot:1999} Tabulation of angle-resolved photoemission (ARPES) data for the
Fermi velocity\cite{Chen:2000} along the zone diagonal in the range $%
0.1<x<0.15$ lead to Fermi velocities in the range range from $1$ to $%
2.5\times 10^{7}cm/s.$ However, recent ARPES data on LSCO\cite{Lanzara:2001}
seem to suggest that the Fermi velocity along the diagonal is doping
independent and of order $2.5\times 10^{7}cm/s$. A weak doping dependence
for the limiting low frequency Fermi velocity is seen in Bi-2212.\cite%
{Johnson:2001} Our {\it bare} Fermi velocity varies linearly with $x$, %
but to compare with ARPES\ experiment, one should include the
effect of residual interactions. The superconducting $d$-wave part of the
interaction does not renormalize the Fermi velocity along the nodal
direction, but antiferromagnetic fluctuations do. For $x=0.15$ and $T=0.05J,$
we find that the bare velocity, estimated from the dispersion of the
spectral weight, renormalizes from $1.27aJ$ to $1.04$ $aJ.$ For $x=0.1,$ the
peak in the spectral function is very broad and in fact the quasiparticle
picture does not strictly applies since $\partial \Sigma /\partial \omega
>0.$ Nevertheless, if we do like in experiment and measure the dispersion of
the spectral weight maximum, we find that for $T$ between\cite{Note-2} $0.2J$
and $0.1J,$ the Fermi velocity renormalizes {\it up} $\left( \text{since }%
\partial \Sigma /\partial \omega >0\right) $from $0.85$ $aJ$ to $0.95\pm
0.05 $ $aJ$. Hence, the physical value of the velocity is much more doping
independent than the bare value suggests. Such doping-independent values
were also found in variational calculations that take into account the
no-double occupancy constraint.\cite{Randeria:2001} For $t/J=3$, and $%
J=125meV,$ our Fermi velocity in physical units is thus about $0.7\times
10^{7}cm/s$, which is smaller than experimental values by a factor $3$ to $%
4. $

\section{Qualitative phase diagram of high $T_{c}$ cuprates}

\label{section5}

Applying the fluctuation approach of the previous section with the effective
theory of the next section, one can use the resulting dynamical SC
susceptibility to estimate the value of the correlation length $\xi _{sc}$
at the temperature where order appears according to mean-field theory. This
length, $\xi _{sc}$, is of order one lattice spacing. We know that quantum
renormalization effects (Kanamori-Brueckner type) should decrease even the
mean-field $T_{c}$ to a lower value $T_{c}^{^{\prime }MF}.$ In addition,
thermal fluctuations decrease it even further. If we look empirically at the
value of $\xi _{sc}$ when one reaches the experimentally determined
superconducting transition temperature $T_{c},$ one finds that at that
temperature, $\xi _{sc}\simeq 3-4$. We know that when the in-plane $\xi _{sc}
$ is sufficiently large, it should induce a three-dimensional SC transition.
Since $\xi _{sc}$ exponentially increases at low temperature in the
renormalized classical regime,\cite{Vilk:1997,Kyung:2001-3} the temperature
where $\xi _{sc}\simeq 3-4$ is not much different from that where it is much
larger. 
\begin{figure}
\centerline{\epsfxsize 8cm \epsffile{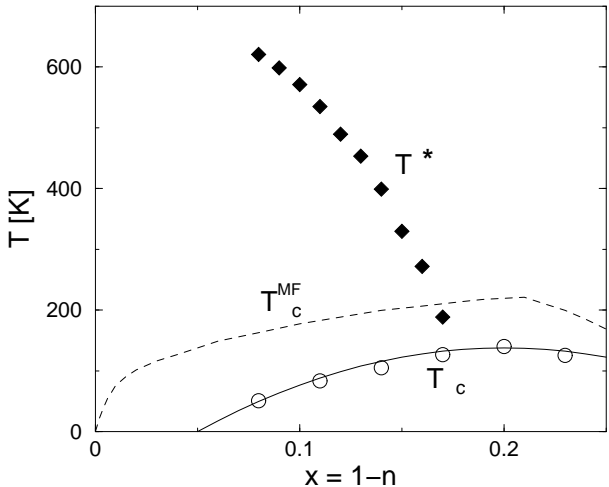}}
\caption{Calculated phase diagram from fluctuation theory.
$T_{c}$ (empty circles) is the SC ordering temperature determined by
fluctuation theory in which the correlation length
$\xi_{sc}$ becomes 4 lattice spacing.
The solid curve is an interpolation of these points
given as $T_{c}=-6122(x-0.20)^{2}+138$ (K).
$T^{MF}_{c}$ (dashed curve) and $T^{*}$ (filled diamonds)
denote mean-field SC ordering temperature and
pseudogap temperature, respectively.}
\label{fig5}
\end{figure}%
The empty circles in Fig.~\ref{fig5} are the value of $T_{c}$ estimated
from the above procedure ($\xi _{sc}=4$). Unlike $T_{c}^{MF}$, $T_{c}$ looks
like a parabola centered at $x_{c}$. The solid curve is an interpolation
formula for the data points given as $T_{c}=-4.22(x-0.20)^{2}+0.095$ (units
of $J$) or $T_{c}=-6122(x-0.20)^{2}+138$ (K). In the same plot, the dashed
curve is the mean-field SC temperature $T_{c}^{MF}$ and the filled diamonds
are the pseudogap temperature $T^{\ast }$. One expects that $O(2)$ SC
fluctuations associated with superfluid stiffness should come into play only
for $T_{c}^{^{\prime }MF}\geq T\geq T_{c}$, as found by Corson {\it et al.} %
\cite{Corson:1999} in their terahertz spectroscopy measurements. This result
contrasts with the SC fluctuation scenario for the pseudogap where SC
fluctuations appear for $T^{\ast }\geq T\geq T_{c}$. The critical region for
SC fluctuations appears larger at underdoping than at overdoping because of
its small superfluid density. A rough estimate for the superfluid density is 
$n/m^{\ast }$ which scales like $x\left( 1-x\right) $ in our approach
because $n\sim \left( 1-x\right) $ and $1/m^{\ast }\sim x$. The pseudogap
temperature $T^{\ast }$ falls from a high value onto the $T_{c}$ line\cite%
{Kyung:2001-4} instead of sharing a common line with $T_{c}$ in the
overdoped region. This is because in our approach the $d$-wave singlet
fluctuations that lead to the normal state pseudogap $T^{\ast }$ are induced
by short-range spin correlations. The latter {\it compete} in the low energy
sector with SC correlations to make $T_{c}$ go to zero near half-filling. %
\cite{Imada:2000} Note however that since $O\left( 2\right) $ SC
fluctuations also cause a pseudogap close to $T_{c}$,\cite{Kyung:2001-3}
some experiments may suggest that the $T^{\ast }$ line continues smoothly
onto the dashed line (or more precisely the $T_{c}^{^{\prime }MF}$ line) 
in the overdoped regime.

In the present study $T_{c}$ and $T_{c}^{MF}$ have their maximum at $%
x_{c}\simeq 0.20$ where condensation energy $U(0)$, mean-field order
parameter $d$ and the coherence gap $\Delta _{cg}$ discussed in the
following section are also maximized. In high $T_{c}$ cuprates, the highest $%
T_{c}$ is at $x=0.16$ slightly lower than the critical doping ($x=0.19$)
where the strength of superconductivity is maximum.\cite{Tallon:2001} This
subtle difference is beyond the scope of the present study.

Although short-range AF correlations persist up to $x_{c}$, they don't lead
to long-range (commensurate) AF order, remaining only as short-range order
even at low temperature, unless the electron density is close to
half-filling. For instance, at $x=0.15$, the AF correlation length $\xi
_{af} $ computed from fluctuation theory saturates at $\xi _{af}\sim 4$ at
the lowest temperature studied while the SC correlation length $\xi _{sc}$
exponentially diverges. Thus the long-range (commensurate) AF phase boundary
is not shown in Fig.~\ref{fig5}. It is expected to lie close to
half-filling where the present formulation is not valid for AF spin
fluctuations since the effective bandwidth, $8xt$ becomes smaller than
interaction strength.\cite{Kyung:2001-4}

\section{The condensate}

\label{sectionSC}

At finite temperature algebraic (Kosterlitz-Thouless) superconducting order
can develop in two dimensions, contrary to AF order, which is prohibited by
the Mermin-Wagner theorem. Since short-range AF correlations can produce a
(dynamical) pseudogap on short-distance scales while thermodynamic
quantities, such as the SC condensate, are calculated in the static and
long-distance limit, we should integrate out AF correlations to study the
low energy physics of the SC state while taking into account the presence of
the AF correlations in an effective manner. In principle, one should be able
to do this and predict the onset of superconductivity from the interacting
Green function of the previous sections, which exhibits a pseudogap.
However, to do this calculation, one needs the corresponding irreducible
vertex. It is not strictly correct to use the Thouless criterion in the form 
$1-JGG=0$ to find the superconducting $T_{c}$ because one cannot use the
bare vertex $J$ to do a calculation with dressed Green functions. Since it
is not known yet\cite{Vilk:1997} how to obtain a reliable approximation for
irreducible vertices in the pseudogap regime, we have to proceed otherwise.
Calculations with constant renormalization to the vertices and bare Green
functions are expected to be more reliable. The phenomenological correlation
length criterion of the previous section is such a calculation.

We want to take into account the presence of short-range AF correlations. To
proceed with a calculation that is done with constant renormalization to the
vertices and bare Green functions, we have to make an additional hypothesis.
We construct an effective low-energy Hamiltonian 
\begin{equation}
H_{eff}=\sum_{\vec{k},\sigma }\varepsilon (\vec{k})c_{\vec{k},\sigma }^{\dag
}c_{\vec{k},\sigma }-V_{eff}\sum_{i}\Delta _{d}^{+}(i)\Delta _{d}(i)\;
\label{eq3-20}
\end{equation}%
such that the effective attraction $V_{eff}$ leads, at $T=0,$ to the same
value of the SC order parameter as that obtained from the full mean-field
theory (Fig.\ref{fig1}(b)). We will use this effective Hamiltonian only to
compute properties of the condensate, including $\xi _{sc}$ of the previous
section. As long as superconductivity, or condensation, is concerned, the
above Hamiltonian already takes into account the effect of (pair breaking)
short-range AF correlations in an effective manner.

\begin{figure}
\centerline{\epsfxsize 8cm \epsffile{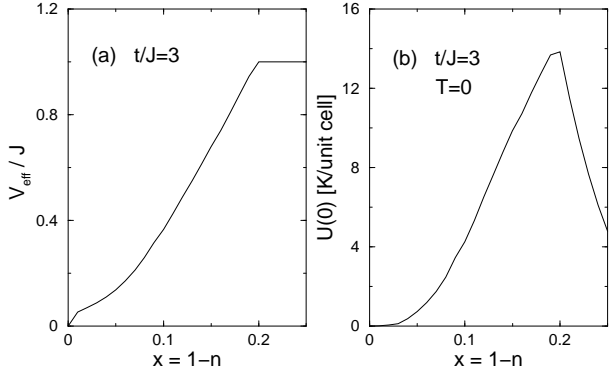}}
\caption{(a) Effective strength of pairing interactions in the
presence of (pair breaking) AF correlations
and (b) calculated condensation energy at $T=0$, as a
function of doping concentration.}
\label{fig2}
\end{figure}%
Above $x_{c}$, $V_{eff}$ is given by $J$, because $m$ vanishes there. We
take the vanishing of $m$ as signaling that the antiferromagnetic
correlation length has become smaller than one lattice spacing. Below $x_{c}$%
, $V_{eff}$ rapidly vanishes with decreasing doping in Fig.\ref{fig2}(a). As
a first application of the above effective Hamiltonian, let us calculate the
condensation energy at $T=0$. The total energies in the SC state and in the
normal state are, in a mean-field approximation, given by 
\begin{eqnarray}
E_{d} &=&\langle \Psi _{d}|H_{eff}-\mu N|\Psi _{d}\rangle  \nonumber \\
&=&\sum_{\vec{k}}[(\varepsilon _{\vec{k}}-\mu )-\frac{(\varepsilon _{\vec{k}%
}-\mu )^{2}}{E_{\vec{k}}}]-|V_{eff}|d^{2}\;,  \nonumber \\
E_{n} &=&\langle \Psi _{n}|H_{eff}-\mu N|\Psi _{n}\rangle  \nonumber \\
&=&\sum_{\vec{k}}[(\varepsilon _{\vec{k}}-\mu )-|(\varepsilon _{\vec{k}}-\mu
)|]\;,  \label{eq3-40}
\end{eqnarray}%
where $E_{\vec{k}}=\sqrt{(\varepsilon _{\vec{k}}-\mu )^{2}+(d\phi _{d}(\vec{k%
})V_{eff})^{2}}$. The $T=0$ condensation energy $U(0)=-(E_{d}-E_{n})$ is
plotted in Fig.~\ref{fig2}(b) as a function of doping concentration with $%
J=125$ meV. $U(0)$ is maximum at $x_{c}$, and rapidly decreases beyond and
below $x_{c}$. This is because $U(0)$ scales like $d^{2}$. This feature is
qualitatively consistent with specific-heat measurements by Loram {\it et al.%
}\cite{Loram:1997} The maximum $U(0)$ in our calculation is about $14K$%
/unit cell at $x_{c}$, two to three times larger than what the above authors
obtained ($5-6K/$unit cell). It is clear that the actual condensation energy
should be smaller than what we calculate since fluctuation effects are
completely neglected in the above calculation. If we had stuck to a
mean-field calculation with both AF and SC order present, we would have
estimated the condensation energy from the difference between the
ground-state energies computed with $d$ finite and with $d$ forced to zero.
For $x>x_{c}$ we would have found the same result as on Fig.~\ref{fig2}(b)
and for $x<x_{c}$ we would have also found a result that decreases with
doping, but it would have been larger than what is found above, suggesting
that the above effective Hamiltonian approach is indeed a better way to take
into account $T=0$ {\it long-range} SC order in the presence of {\it %
short-range} AF correlations.

Another quantity of interest is the coherence-gap $\Delta _{cg}.$ Deutscher %
\cite{Deutscher:1999} has proposed that this gap, accessible through Andreev
reflection experiments with point-contact spectroscopy, can be different
from the single-particle gap observed in tunneling. Recent experiments\cite%
{Gonnelli:2001} seem to confirm this. To estimate this coherence gap, one
can use the mean-field solution of the effective BCS model Eq.(\ref{eq3-20}%
). However, a better estimate can be obtained by computing the precursor
pseudogap with the method explained below. Indeed, it is known\cite%
{Kyung:2001-3} in the attractive Hubbard model that this precursor pseudogap
reaches its asymptotic low-temperature value very rapidly. In addition, this
precursor pseudogap contains quantum renormalization effects that are absent
from a pure mean-field calculation. The following calculation is in the
spirit of Ref.\cite{Kyung:2001-4}. First we compute how much the $d$-wave
pair correlation function $\chi _{pp}$ is enhanced over $\chi _{pp}^{0}$ by
applying the local-pair sum rule for $\chi _{pp}$ 
\begin{equation}
\frac{T}{N}\sum_{q}\chi _{pp}(q)e^{-i\nu _{m}0^{-}}=\langle |\Delta
_{d}(0)|^{2}\rangle \;.  \label{eq4-10}
\end{equation}%
The right-hand side of the above equation is evaluated in the mean-field
state of Eq.~\ref{eq3-20} and $\chi _{pp}$ is related to $\chi _{pp}^{0}$
through the renormalized vertex $V_{pp}$ 
\[
\chi _{pp}(q)=\frac{\chi _{pp}^{0}(q)}{1-V_{pp}\chi _{pp}^{0}(q)}\;, 
\]%
where the irreducible susceptibility is defined as 
\[
\chi _{pp}^{0}(q)=\frac{T}{4N}\sum_{k}(\phi _{d}(\vec{k})+\phi _{d}(\vec{q}-%
\vec{k}))^{2}G^{0}(q-k)G^{0}(k)\;. 
\]%
Finally the following self-energy is used to estimate the effect of $d$-wave
pairing correlations on quasiparticles. 
\begin{eqnarray}
\Sigma _{pp}(k) &=&-\frac{1}{4}V_{eff}V_{pp}\frac{T}{N}\sum_{q}  \nonumber \\
&&(\phi _{d}(\vec{k})+\phi _{d}(\vec{q}-\vec{k}))^{2}\chi
_{pp}(q)G^{0}(q-k)\;.  \label{eq4-40}
\end{eqnarray}%
This procedure is analogous to that used earlier in the case of the
attractive Hubbard model ($s$-wave symmetry)\cite{Kyung:2001-3} to obtain
good agreement with QMC calculations. Even though the above formula has not
been obtained with the same rigor, one expects it to give a good estimate
of what happens in the case of $d$-wave symmetry.

\begin{figure}
\centerline{\epsfxsize 8cm \epsffile{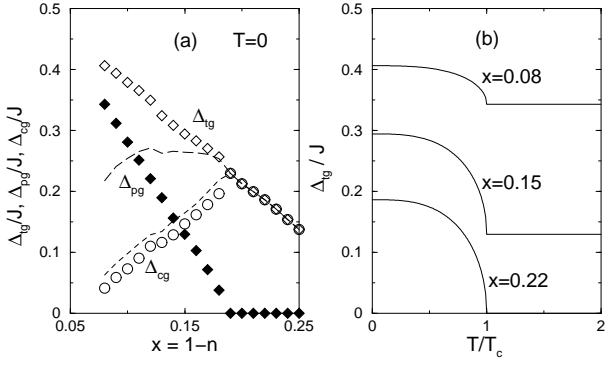}}
\caption{(a) Calculated total excitation gap $\Delta_{tg}$ (empty diamonds),
pseudogap $\Delta_{pg}$ (filled diamonds)
and condensation gap $\Delta_{cg}$ (empty circles)
at $T=0$ as a function of doping.
The dashed and long-dashed curves denote
$\Delta_{tg}-\Delta_{pg}$ and
$\sqrt{\Delta^{2}_{tg}-\Delta^{2}_{pg}}$, respectively.
(b) The total excitation gap $\Delta_{tg}$ as a function
of temperature for three representative doping levels
($x=0.08$ for underdoping, $x=0.15$ for optimal doping and
$x=0.22$ for overdoping).}
\label{fig4}
\end{figure}%
Figure~\ref{fig4}(a) shows $\Delta _{cg}$ (empty circles) together with $%
\Delta _{tg}$ (empty diamonds) and $\Delta _{pg}$ (filled diamonds) as a
function of doping at $T=0$. Below $x_{c},$ $\Delta _{cg}$ monotonically
increases with doping and above $x_{c}$ it reduces to $\Delta _{tg}$. The
doping dependence of $\Delta _{cg}$ is qualitatively similar to $T_{c}^{MF}$
and $d,$ as found in the experiments of Ref.\cite{Gonnelli:2001}. It is
surprising to observe that $\Delta _{cg}$, $\Delta _{tg}$ and $\Delta _{pg}$
are more or less linear in doping concentration, in spite of the fact that
they are obtained through highly nonlinear equations. In fact the linear
behavior of $\Delta _{tg}$ and $\Delta _{pg}$ in doping are observed by
Tallon and Loram \cite{Tallon:2001}, and by Ding {\it et al.}\cite{Ding:2000}
in their respective experiments. To find out an empirical relation between $%
\Delta _{cg}$, $\Delta _{tg}$ and $\Delta _{pg}$, we plotted $\sqrt{\Delta
_{tg}^{2}-\Delta _{pg}^{2}}$ (long-dashed curve) and $\Delta _{tg}-\Delta
_{pg}$ (dashed curve). The former is nearly independent of doping, in
agreement with the result often found by other authors using $\Delta _{cg}=%
\sqrt{\Delta _{tg}^{2}-\Delta _{pg}^{2}}$\cite{Chen:1998} to extract the
coherence gap from experiment. Within our numerical uncertainties, the
coherence-gap $\Delta _{cg}$ is approximately given as $\Delta _{tg}-\Delta
_{pg}$ and its behavior is easier to reconcile with the doping dependence of 
$T_{c}$ and $d$. Recently Krasnov\cite{Krasnov:2002} also found that the
coherence gap (dubbed as the SC gap in his paper) shows a doping dependence
similar to that of $T_{c}$.

When $\Delta _{tg}\simeq \Delta _{pg}+\Delta _{cg}$ is generalized to finite
temperature, the total excitation gap is composed of the normal state
pseudogap and the coherence-gap in Fig.\ref{fig4}(b). $\Delta _{pg}$ is
virtually temperature independent from $T^{\ast }$ to $T=0$, because it is
caused by short-range spin correlations of order one lattice spacing. On the
other hand, we expect that $\Delta _{cg}$ is strongly temperature dependent
since it opens up at $T_{c}.$ For purposes of illustration, we added a
BCS-like temperature dependence to $\Delta _{cg}.$ Its zero
temperature-value and $T_{c}$ were estimated above. As one can see from \ref%
{fig4}(b), at large underdoping ($x=0.08$) $\Delta _{tg}$ (or ARPES leading
edge gap) is mainly given by $\Delta _{pg}$, while at optimal doping ($%
x=0.15 $) it is more or less equally made of $\Delta _{pg}$ and $\Delta
_{cg} $ below $T_{c}$. At large overdoping ($x=0.22$) $\Delta _{tg}=\Delta
_{cg}$, thus leading to the conventional BCS behavior. This feature is also
qualitatively consistent with ARPES experiments.\cite{Ding:1996,Loeser:1996}

A similar behavior of $\Delta _{tg}$ is used by Chen {\it et al.}\cite%
{Chen:1998} to calculate the temperature dependence of some physical
quantities within their SC fluctuation theory of the pseudogap. The crucial
difference between their theory and ours is that their $\Delta _{pg}$
vanishes at $T=0$ so that when superconductivity is completely destroyed,
for instance, by a strong magnetic field, their ground state becomes a
perfect Fermi liquid. In our picture, however, we predict that the ground
state has a pseudogap.\cite{Kyung:2001-4} Our calculated temperature
dependence of the total excitation gap (or ARPES leading edge gap from the
experimental point of view) contrasts with some theories based on the idea
of spin-charge separation. In these theories the only difference between $A(%
\vec{k},\omega )$ above and below $T_{c}$ is the line shape due to the Bose
condensation of holons, namely, the appearance of the quasiparticle peak
below $T_{c}$ without change of the gap magnitude itself. At large
underdoping, there is indeed not much change in the ARPES leading edge gap %
\cite{Ding:1996,Loeser:1996}, because from our point of view, $\Delta _{tg}$
is composed mainly of the temperature independent $\Delta _{pg}$. At optimal
doping the above ARPES experiments show a significant change of the
excitation gap itself with decreasing temperature in addition to the
sharpening of the peak. From our point of view, this significant change of $%
\Delta _{tg}$ (at optimal doping) with temperature is due to the strong
temperature dependence of the coherence gap $\Delta _{cg}$ near the
transition.

Since the zero-temperature critical field $H_{c2}$ is a property of the
condensate, the above $V_{eff}$ model can also be used to estimate it. Using
the $d$-wave estimate\cite{Kim:1998} $H_{c2}=0.521\Phi _{0}\left( 2\pi \xi
_{0}^{2}\right) ^{-1}$ with $\xi _{0}=\hbar v_{F}\left( \pi \Delta
_{cg}\right) ^{-1}$ and $\Phi _{0}=hc\left( 2e\right) ^{-1}$, we find that $%
H_{c2}\left( x=0.08\right) =\allowbreak 42T$ and $H_{c2}\left( x=0.15\right)
=151T$. The experimental values\cite{Ando:1999} are, respectively, around $%
30T$ and $60T$, somewhat smaller. However, there is still disagreement on
the experimental value of $H_{c2}$ and recent estimates are larger\cite%
{Luo:2001}. The above model is extremely crude since it does not take into
account the anisotropy and interaction renormalization of the Fermi
velocity. Within this crude model, we have that $H_{c2}$ will decrease at
dopings larger than optimal because above optimal doping $\Delta _{cg}$
decreases.

\section{Discussion}

\label{section6}

In this Section we compare the present results with the predictions of some
of the leading theories for the high $T_{c}$ superconductors and we provide
further comparison with experiments. Before we proceed, we should stress
that the starting point of our approach is meant to apply to the
intermediate to strong-coupling regime and thus should be distinguished from
spin-fluctuations theories that apply from weak to intermediate coupling.%
\cite{Chubukov:2002,Kyung:2002}

First, consider the question of why the superconducting $T_{c}$ decreases
near half-filling. As is clear by now, in our study the reduction of $T_{c}$
near half-filling occurs already {\it at mean-field level} due to the
competition with AF correlations (or local moment formation) that open up a
mean-field gap, making SC order less favorable. Although SC fluctuations
will indeed make $T_{c}$ decrease with decreasing doping because of the
smaller superfluid density\cite{Emery:1995,Chen:1998}, such fluctuations are
not necessary to have the correct qualitative behavior.

Second, what is the nature of the normal state pseudogap? In this paper the
pseudogap is just a crossover phenomenon due to short-range {\it equal-time}
spin correlations which induce short-range {\it dynamical }fluctuations in
the particle-particle channel. The latter are very effective in creating a
pseudogap for several reasons. First note that the mass renormalization
makes the bandwidth and the Fermi velocity quite small. In addition, the
pseudogap opens up near the band edges where the local Fermi velocity is
smallest. This means that the correlation length associated with the
fluctuations does not need to be large to be in the renormalized classical
regime. Second, for $q=0$ $d$-wave singlet fluctuations, no umklapp condition needs to
be satisfied contrary to antiferromagnetic hot spots.\cite{Yury:1997} The
proposed origin for the pseudogap contrasts with some of the recent
proposals in which {\it mean-field} flux phase,\cite{Wen:1996} circulating
current phase,\cite{Varma:1999} and $d$-density wave phase \cite%
{Chakravarty:2001} are interpreted as the pseudogap state. In these
scenarios, the pseudogap state has broken symmetry such as time reversal,
translational and rotational symmetries, and the crossover-like behavior
observed in experiments is argued to be caused by an impurity effect.

Third, let us consider experimental evidence that there are significant
short-range AF correlations below $x_{c}$. Recent muon spin relaxation\cite%
{Sonier:2000,Panagopoulos:2001} and ac-susceptibility measurements, show the
existence of slow spin fluctuations below $x=0.19$. By using polarized and
unpolarized elastic neutron scattering experiments as well as zero-field
muon spin resonance, Sidis {\it et al.} \cite{Sidis:2001} observed an
unusual commensurate AF phase on a nanosecond time scale that coexists with
superconductivity in underdoped YBa$_{2}$Cu$_{3}$O$_{6.5}$. Lake {\it et al.}
\cite{Lake:2001} found AF correlations inside a vortex core in underdoped
cuprates. Most recently, Hodges {\it et al.}\cite{Hodges:2001} have even
shown that the addition of $Co$ to an optimally doped $YBCO$ compound
induces antiferromagnetism above $T_{c}$ that survives and coexists with $%
d-wave$ superconductivity below $T_{c}.$ The coexistence seems to be at the
microscopic level with $\xi >200$\AA $.${\it \ }In the present paper, the
presence of AF correlations below $x_{c}$ does not come from any {\it extra}
symmetry between AF and SC correlations such as $SO(5)$ symmetry, but from
the fact that the $J$ part of the $t-J$ Hamiltonian has those competing
correlations in the low energy sector. Table 1 gives a few results for the
temperature-dependent antiferromagnetic correlation length in our approach
for two dopings. As expected, the correlations decrease with doping. An
upper bound, order of magnitude estimate for the value of the static local
moment is given by the value of $m$ in Fig.\ref{fig1}(b)

%TCIMACRO{\TeXButton{Begin table}{\begin{table}} }%
%BeginExpansion
\begin{table}
%EndExpansion
%TCIMACRO{
%\TeXButton{Caption}{\caption{Temperature dependent antiferromagnetic 
%correlation length evaluated for two dopings using fluctuation theory. The fluctuations
%may either be commensurate (C) or incommensurate (IC). 
%The mesh used in the calculation was $128 \times 128$
%hence the largest number in this table is a rough estimate.}
%}}%
%BeginExpansion
\caption{Temperature dependent antiferromagnetic 
correlation length evaluated for two dopings using fluctuation theory. The fluctuations
may either be commensurate (C) or incommensurate (IC). 
The mesh used in the calculation was $128 \times 128$
hence the largest number in this table is a rough estimate.}
%
%EndExpansion

\begin{tabular}{ccc}
& $x=0.1$ & $x=0.15$ \\ 
$T=0.02J$ & $133$ (IC) & $4.1$ (IC) \\ 
$T=0.04J$ & $25.5$ (C) & $2.6$ (IC) \\ 
$T=0.06J$ & $6.1$ (C) & $1.9$ (IC)%
\end{tabular}

%TCIMACRO{\TeXButton{End table}{\end{table}}}%
%BeginExpansion
\end{table}%
%EndExpansion

Fourth, we point out that there is numerical and experimental evidence that
short-range spin correlations are closely related to the normal state
pseudogap. In their recent calculations obtained from the {\it dynamical}
cluster approximation (DCA) for the Hubbard model, Jarrell {\em et al.}\cite%
{Jarrell:2000} noted that the fall of $T^{\ast }$ with doping is closely
tied to the diminishing of the short-range spin correlations. In the exact
diagonalization study of the $t-J$ model, Sakai and Takahashi\cite%
{Sakai:2000} found that the pseudogap behavior is associated with the
development of static AF spin correlations with decreasing temperature. In
his high temperature series expansion study of the $t-J$ model, Putikka\cite%
{Putikka:2000} showed that pseudogap crossover occurs when static AF
correlation length is about one lattice spacing. By analyzing various
experiments (Raman, spin-lattice relaxation rate, ARPES, Zn-substitution
effects, inelastic neutron scattering experiment and so on), Tallon and Loram%
\cite{Tallon:2001} concluded that the pseudogap is intimately connected with
short-range AF correlations. Short-range dynamical AF correlations can also
cause the normal state pseudogap through AF spin fluctuations. As explained
above however, it was found in a recent paper\cite{Kyung:2001-4} that the
pseudogap always appears earlier in the dynamical particle-particle channel
than in the dynamical AF spin fluctuation channel, even though the
short-range equal-time correlations that generate these dynamical
fluctuations are due to AF correlations. One of the difficulties with AF
spin fluctuation scenario of the pseudogap is how the normal state pseudogap
happens to have the same momentum dependence and magnitude as the SC gap
near $T_{c}$,\cite{Schmalian:1998} as ARPES and tunneling experiments show.
In our case, this occurs naturally.
Our results remain qualitatively the same when $t^{\prime }$ is $-0.3t.$

Fifth, is $x_{c}\simeq 0.19-0.20$ accidental for the special choice of $%
t/J=3 $? According to the calculations of the slave-boson mean-field
solution of the Hubbard model\cite{Kotliar:1986,Meyer:1993}, the onset of
short range AF correlations starts to appear from $x=0.20-0.21$ for a wide
range of $U=7t-16t$ (See Fig.1 of Ref.~\cite{Meyer:1993} for more details).
In our approach, $x_{c}$ varies from $0.22$ to $0.17$ when $t/J$ varies from 
$2.5$ to $4.$ It is believed that a realistic strength of the Coulomb
repulsion $U=4t/J$ is of the order of the bare bandwidth $8t$ in two
dimension.\cite{Aeppli:2001}

Last, through the microscopic separation of hole-rich SC regions from AF
regions,\cite{Zaanen:1989} the stripe structure tends to maintain AF
correlations more effectively than the other case in which they are
uniformly suppressed by doped holes. Then one can surmise that in the stripe
state $T_{c}$ is somewhat suppressed from that in the uniform state. While
the differences between the stripe state and the uniform state can be quite
substantial, we do not see the stripes as necessary to obtain
superconductivity. The main results reached in this paper are not expected
to qualitatively change in the presence of a dynamical stripe structure.

\section{Conclusion}

\label{section7}

In this paper we proposed a simple phenomenological procedure that allows
one to study the competition between antiferromagnetism and $d$-wave
superconductivity in the high $T_{c}$ cuprates. The no-double-occupancy
constraint is taken into account in the effective mass approximation.
Correlation functions that are local in space and time are evaluated from a
mean-field factorization of the $t-J$ Hamiltonian in both the AF and the $d-$%
wave SC channels. (The $d-$density wave channel does not contribute in the
whole doping range.) The local correlation functions are then used in the
fluctuation-dissipation theorem to compute renormalized vertices that allow
one to obtain the full {\it dynamical} susceptibilities. In other words, in
this approach, equal-time correlation functions determine the value of
effective vertices in all available channels. In particular, short-range
equal-time AF correlations determine the value of effective interactions in
the particle-particle channel. This effective weak to intermediate-coupling
approach cannot work for $x<0.08$ where the bandwidth becomes less than the
interaction strength.

The calculated total excitation gap $\Delta _{tg}$ (in the superconducting
state) and the normal state pseudogap $\Delta _{pg}$ are in good agreement
with experimental results, as shown earlier in Ref.\cite{Kyung:2001-4}.
Obtaining a description of superconducting properties arising from a highly
correlated state remains a challenge. Here we have taken the simplest
approach. We compute an effective strength of pairing interactions which
takes into account (pair breaking) AF correlations, $V_{eff}$, by requiring
that the zero-temperature order parameter $d$ obtained with $V_{eff}$ equals
that obtained in the full mean field equations corresponding to Eq.(\ref%
{eq2-40}). This effective interaction allows one to obtain properties of the
condensate (and of the condensate only), namely the condensation energy $U(0)
$, the coherence gap $\Delta _{cg}$ that has been observed in Andreev
reflection \cite{Gonnelli:2001} as well as $H_{c2}$. The calculated
coherence gap closely follows the doping dependence of $T_{c}$ or $d$, and
is approximately given as $\Delta _{cg}\sim \Delta _{tg}-\Delta _{pg}$
within our numerical uncertainties. A {\it qualitative} phase diagram for
the cuprates may thus be obtained. The systematic decrease of $U(0)$, $%
\Delta _{cg}$ and $H_{c2}$ with decreasing doping below $x_{c}\sim 0.2$ can
be understood as a result of the competition between AF and SC order
occurring in the low energy sector of the mean-field $t-J$ Hamiltonian. On
the other hand, in the present picture the systematic increase of $\Delta
_{pg}$ and $\Delta _{tg}$ when $x$ decreases below $x_{c}\sim 0.2$ is due to
the growing short-range {\it equal-time} spin correlations which induce {\it %
dynamical} singlet fluctuations with $d$-wave symmetry, low characteristic
frequency but with small correlation length. As pointed out before in Ref. %
\cite{Kyung:2001-4} one prediction of this approach is that the pseudogap
should survive if the superconducting state is destroyed by a magnetic
field. This is consistent with the observation of antiferromagnetic ordering
inside vortex cores.\cite{Lake:2001} Neutron scattering experiments should
be able to check from energy integrated structure factor that the
short-range spin correlation functions are in good agreement with the
mean-field predictions. Finally, we expect $H_{c2}$ to decrease with doping
in the overdoped region.

Although some of the results of this paper are in quantitative agreement
with experiment, the overall description is definitely qualitative only. The
main disagreement with experiment is that the {\it renormalized} Fermi
velocity that we find is three to four times smaller than measured. A
satisfactory theory of high $T_{c}$ will necessitate a better treatment of
strong-coupling short-range correlations.

\acknowledgements

We are grateful to H. Ding for sending his data and to P. Fournier and L.
Taillefer for numerous discussions and for pointing out some references. We
also thank J. Carbotte and K. Maki for useful correspondence on $d-$wave
superconductivity. The present work was supported by a grant from the
Natural Sciences and Engineering Research Council (NSERC) of Canada, the
Fonds pour la formation de Chercheurs et l'Aide \`{a} la Recherche (FCAR) of
the Qu\'{e}bec government and the Tier I Canada Research Chair Program
(A.-M.S.T.).


\begin{references}
\bibitem{NoteDefSO(5)} Our definitions of AF and SC order parameters are
equivalent to those appearing in SO(5) theory.
\bibitem{Note-1} Our approach is invalid for $x<0.08$ because $8tx<2J$ in
that doping range, invalidating the weak coupling approach used in the
calculation of dynamical quantities. For $x$ larger than about $0.4,$ both
zero temperature mean-field order parameters vanish meaning that
correlations become more subtle. There is also no reason to believe that the
effective-mass approximation is valid far from half-filling.
\bibitem{NoteApprox} This good agreement is in part fortuitous considering
the various approximations used in this work. These include replacing the
no-double-occupancy projector by $x$, evaluating local correlation functions
in the mean-field state.
\end{references}
\end{document}